\documentclass[aps,nofootinbib,preprint]{revtex4}
\usepackage{amsfonts,amssymb,amsmath}
\usepackage{graphicx}
\usepackage{subfigure}
\usepackage{wrapft}
\def\d{\delta}
\def\g{\sqrt{-g}}
\def\nt{\notag}
\newcommand{\f}[2]{\frac{#1}{#2}}
\newcommand{\mk}[1]{\left( #1 \right)}
\newcommand{\kk}[1]{\left[ #1 \right]}

\newcommand{\be}{\begin{equation}}
\newcommand{\ee}{\end{equation}}
\newcommand{\DE}{{\rm DE}}
\newcommand{\lcdm}{{\rm \Lambda CDM}}
\newcommand{\frg}{{\rm fRG}}
\def\eff{{\rm eff}}
\def\eV{{\rm eV}}

\def\Mpcin{{\rm Mpc}^{-1}}
\def\bfk{{\bf k}}
\begin{document}

\title{
Matter Power Spectrum in $f(R)$ Gravity with Massive Neutrinos
}

\author{
Hayato Motohashi$^{~1,2}$, Alexei A. Starobinsky$^{~2,3}$,
and Jun'ichi Yokoyama$^{~2,4}$
}

\address{
$^{1}$ Department of Physics, Graduate School of Science,
The University of Tokyo, Tokyo 113-0033, Japan \\
$^{2}$ Research Center for the Early Universe (RESCEU),
Graduate School of Science, The University of Tokyo, Tokyo 113-0033, Japan \\
$^{3}$ L. D. Landau Institute for Theoretical Physics,
Moscow 119334, Russia \\
$^{4}$ Institute for the Physics and Mathematics of the Universe(IPMU),
The University of Tokyo, Kashiwa, Chiba, 277-8568, Japan
}

\begin{abstract}
The effect of massive neutrinos on matter power spectrum is
discussed in the context of $f(R)$ gravity. It is shown that the
anomalous growth of density fluctuations on small scales due to
the scalaron force can be compensated by free streaming of
neutrinos. As a result, models which predict observable deviation
of the equation-of-state parameter $w_\DE$ from $w_\DE=-1$ can be
reconciled with observations of matter clustering if the total
neutrino mass is $O(0.5~\eV)$.
\end{abstract}

\begin{flushright}
RESCEU-9/10
\end{flushright}

\maketitle


$f(R)$ gravity is a simple and nontrivial extension of General
Relativity (GR) which has recently received much attention as
an alternative to the ${\rm \Lambda}$-Cold-Dark-Matter ($\lcdm$) model
that can explain the present cosmic acceleration. 
The basic idea is to modify the Einstein-Hilbert action 
by using the general function of Ricci scalar $R$ as 
\be \label{act} S=\int d^4x \g f(R)+S_m, \ee 
where $S_m$ indicates a matter action
(see Ref.~\cite{FT10} for a recent review and
an extensive list of references). If $f''(R)$ is not zero
identically, this modified gravity contains an additional degree
of freedom corresponding to a new scalar field, dubbed scalaron in
Ref.~\cite{S80} where the simplest, $f(R)=R+R^2/6M^2$, variant of such
theory (with small one-loop quantum gravitational corrections) was
used to construct an internally self-consistent inflationary model
of the early Universe with a graceful exit from inflation to the
subsequent radiation-dominated Friedmann-Robertson-Walker (FRW)
stage after the gravitational creation of matter and reheating. 
However, $f(R)$ gravity can be used to describe dark energy in the present
Universe as well, although with the use of more complicated functions $f(R)$
having a nontrivial structure for low values of $R$.

To have the correct Newtonian limit for $R\gg R_0\equiv
R(t_0)\sim H_0^2$, where $t_0$ is the present moment and $H_0$ is
the Hubble constant, as well as the standard matter-dominated FRW
stage with the scale factor behaviour $a(t)\propto t^{2/3}$ driven
by cold dark matter and baryons, the following conditions should
be fulfilled: \be |f(R)-R|\ll R,~~|f'(R)-1|\ll 1,~~Rf''(R)\ll 1,
~~R\gg R_0~, \label{ineq} \ee where the prime denotes a derivative
with respect to the argument $R$. The same conditions guarantee
smallness of non-GR corrections to a space-time metric for more
general backgrounds for which GR has to be used in full. In
addition, the stability condition $f''(R)>0$ has to be satisfied
to guarantee that the standard matter- and radiation-dominated FRW
stages remain attractors with respect to an open set of
neighboring isotropic cosmological solutions in $f(R)$ gravity. 
In quantum language, this condition means that the scalaron is not a
tachyon. Note that the other stability condition, $f'(R)>0$, which
means that gravity is attractive and the graviton is not a ghost,
is automatically fulfilled in this regime. Specific functional
forms that satisfy these and other necessary conditions have been
proposed in Refs.~\cite{Hu:2007nk,AB07,Starobinsky:2007hu},
and much work has been conducted on their cosmological consequences.

We can express field equations derived from action \eqref{act} in the
following Einsteinian form,
\be
 R^{\mu}_{\nu}-\frac{1}{2}\delta^{\mu}_{\nu}R=
-8\pi G\mk{T^{\mu}_{\nu (m)}+T^\mu_{\nu (\DE)}}, \ee
where
\be \label{EMtensor} 8\pi G T^\mu_{\nu (\DE)}\equiv
(F-1)R^\mu_\nu-\f{1}{2}(f-R)\delta^{\mu}_{\nu}
+(\nabla^\mu\nabla_\nu-\delta^{\mu}_{\nu}\square)F,\ee
and we have also defined $F(R)\equiv f'(R)=df/dR$.
Working in the
spatially flat Friedmann-Robertson-Walker (FRW) space-time with
the scale factor $a(t)$, we find
\begin{align}
3H^2&=8\pi G\rho-3(F-1)H^2+\frac{1}{2}(FR-f)-3H\dot{F},\label{hubble}\\
2\dot{H}&=-8\pi G\rho -2(F-1)\dot{H}-\ddot{F}+H\dot{F},\label{hdot}
\end{align}
where $H$ is the Hubble parameter and $\rho$ is the energy density
of the material content, which we assume to consist of
nonrelativistic matter.
From these expressions, the effective DE equation-of-state parameter
$w_\DE\equiv P_\DE/\rho_\DE$ is given by
\begin{align}
1+w_\DE &= \f{-2\dot{H}-8\pi G\rho}{3H^2-8\pi G\rho} \nt\\
\label{wDE} &= \f{2(1-F)(-\ddot{a}/a+H^2)+\ddot{F}-H\dot{F}}{3(1-F)
(-\ddot{a}/a)+(R-f)/2-3H\dot{F}}.
\end{align}
Note that the phantom crossing is naturally realized in viable
$f(R)$ theories as shown in Refs.~\cite{Hu:2007nk,ABS09,AGSS10,Motohashi:2010tb}.

We specifically adopt the following functional form that 
satisfies all the above requirements\cite{Starobinsky:2007hu}, 
\be \label{fR} f(R)=R+\lambda R_s\kk{\mk{1+\f{R^2}{R_s^2}}^{-n}-1}, \ee 
where $n,~\lambda$, and $R_s$ are model parameters. It is
sufficient to describe the recent evolution of the
Universe, although it has to be modified both for $R<R_0$ (including
the region $R<0$) and for very large positive $R$ to avoid
problems in the early Universe cosmology
(see Refs.~\cite{Motohashi:2009qn,ABS09}).

In $f(R)$ gravity, the evolution equation for density fluctuation
with comoving wave number $k$ in Fourier space, $\d_\bfk$, in the
deeply subhorizon regime is given by \cite{Zhang:2005vt,Tsujikawa:2007gd} 
\be \label{dedp} \ddot \d_\bfk + 2H\dot \d_\bfk - 4\pi G_\eff \rho \d_\bfk = 0, \ee 
where
\be \label{Geff} G_\eff=\f{G}{F} \f{1+4\f{k^2}{a^2}\f{F'}{F}}
{1+3\f{k^2}{a^2}\f{F'}{F}}. \ee 
In the high-curvature regime, the
differential equation \eqref{dedp} can be solved analytically
\cite{Motohashi:2009qn}, and we analyzed it in the general regime
in terms of numerical calculation in the previous paper
\cite{Motohashi:2010tb}. It is revealed that the effective
gravitational coefficient $G_\eff$ enhances the growth of density
fluctuations compared with that in the standard $\lcdm$ model due
to an extra force mediated by the scalaron. To suppress
the deviation of the matter power spectrum from that in the
$\lcdm$ model within, say $10\%$, the parameter space for $n$ and
$\lambda$ should be restricted. For example, $\lambda$ should be greater
than 8 in $n=2$ \cite{Motohashi:2010tb}.

In this paper, we show that the anomalous growth of density
fluctuations on small scales due to the scalaron is rectified if
neutrino masses are taken into consideration. That is, the free
streaming of neutrinos partially erases small-scale density
fluctuations, which can be harmful in Einstein gravity, to
compensate their unwanted extra growth in $f(R)$ gravity.

To set up the initial condition,
we calculate the evolution of density fluctuations
for $z \geq 10$ using CAMB
\cite{Lewis:1999bs,CAMB.info},
and obtained $P(k,z=10)$, power spectrum at $z=10$.
We assume one mass eigenstate for neutrino,
and all three neutrinos have practically the same masses.
Note that the density parameter of neutrinos is
\be
\Omega_\nu h^2\simeq
\left\{
\begin{array}{ll}
1.685\times 10^{-5} &\quad \mk{\sum m_\nu=0} \\
\displaystyle \f{\sum m_\nu}{94.1~\eV} &\quad \mk{\sum m_\nu
\not=0},
\end{array}
\right. \ee 
and the free streaming of massive neutrinos suppresses
fluctuations below the scale \cite{Hu:1997vi,Eisenstein:1997jh}
\be \label{neu} k_{\rm fs}(z) \simeq \f{0.35}{(1+z)^{1/2}}
\mk{\f{m_\nu}{1~\eV}} \mk{\f{\Omega_m}{0.27}}^{1/2}
h~\Mpcin. \ee 
Therefore, $\d_\bfk$ in Eq.~\eqref{dedp} implies density fluctuations for CDM below that scale.

At $z\gtrsim 10$, the effect of modified gravity is sufficiently small.
Therefore, we start to solve the evolution of density fluctuations in $f(R)$
gravity using Eq.~\eqref{dedp} at $z=10$. 
We have found that the
physical wave number crossing the scalaron mass today is given by
\be \label{pwcsmt} \f{k_s}{a_0} \simeq 3.2~\lambda^{1.88}H_0, \ee
for $n=2$\cite{Motohashi:2010tb}, which is to be compared with
Eq.~\eqref{neu}. 
It takes $1.07\times 10^{-3} h$ and
$8.44\times 10^{-3} h~\Mpcin$ for $\lambda=1$ and $3$,
respectively.

\begin{figure}[t]
\centering
\subfigure{\includegraphics[width=60.3mm]{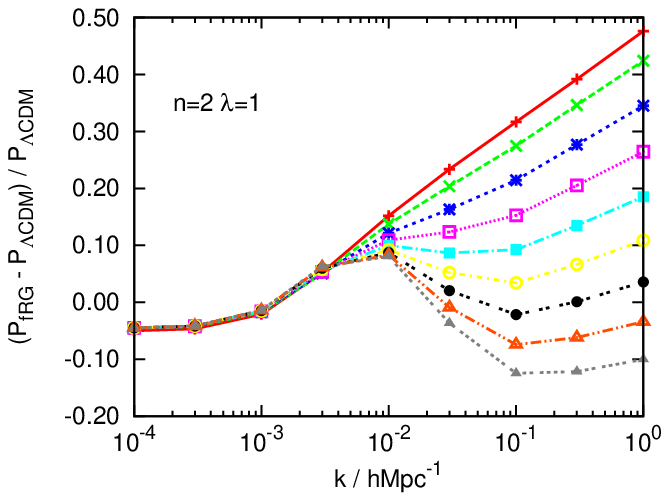}}
\subfigure{\includegraphics[width=78.3mm]{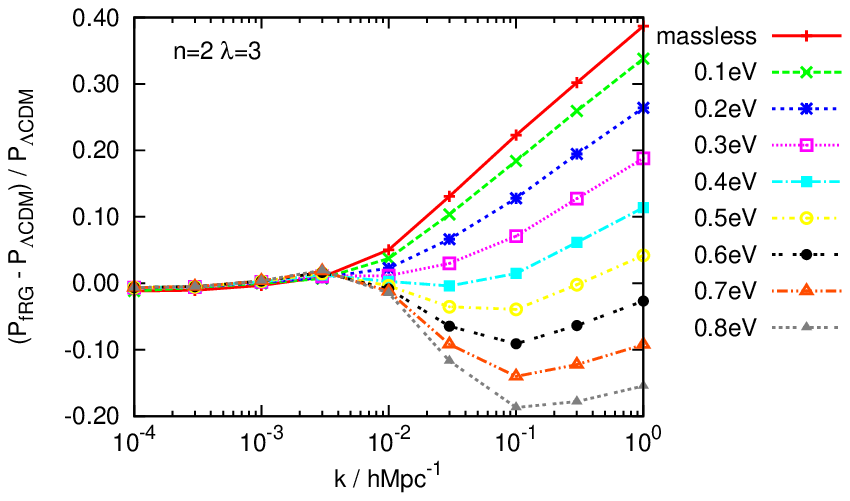}}
\caption{Deviation of power spectrum for $n=2,~ \lambda=1,~3$.
 Captions indicate total neutrino mass, and
 it is assumed that three neutrinos have the same masses.}
\label{fRG_LCDM_mn_n2l13}
\end{figure}

The present power spectrum is constructed as 
\be P(k,z=0)=P(k,z=10)\mk{\f{\d(z=0)}{\d(z=10)}}^2. \ee 
Figure \ref{fRG_LCDM_mn_n2l13} shows cancellation of the contributions
from $f(R)$ gravity and neutrino masses. Equations \eqref{neu} and
\eqref{pwcsmt} explain the threshold wave number. In particular,
the transition occurs at a similar wave number for $\lambda=3$ when we
take $n=2$. The deviation at $k=0.174h~{\rm Mpc}^{-1}$, which is
the scale corresponding to $\sigma_8$ normalization
\cite{Motohashi:2010tb}, is minimum when the total neutrino masses are
$0.6$ and $0.5~\eV$ for $\lambda=1$ and $3$, respectively.

\begin{figure}[t]
\centering
\subfigure{\includegraphics[width=70.4mm]{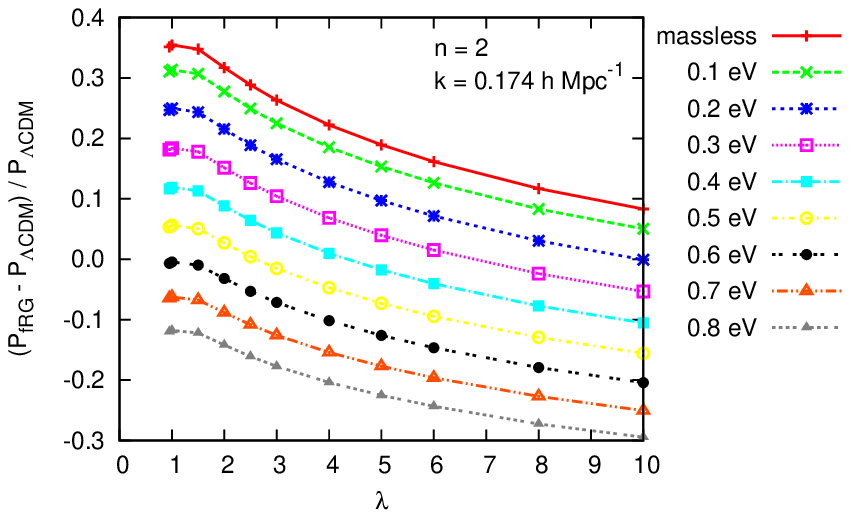}}
\subfigure{\includegraphics[width=56mm]{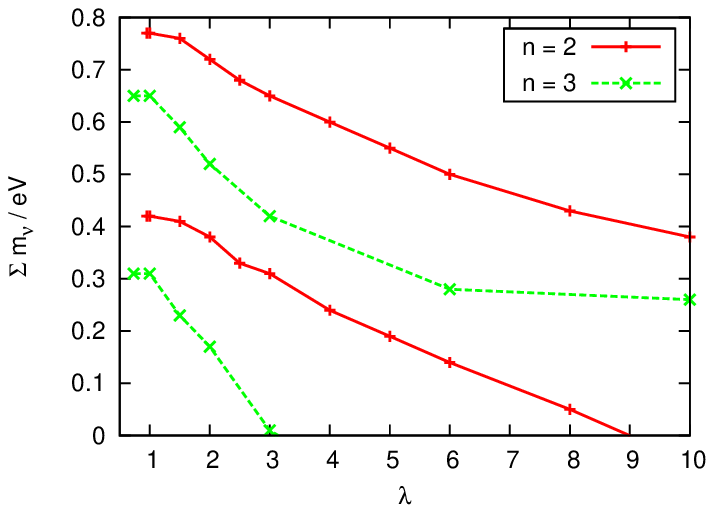}}
\caption{ Left: Relative deviation of linear matter power spectrum
from $\lcdm$ model at $k=0.174h~\Mpcin$ as a function of $\lambda$
for various values of neutrino mass. Right: The region between two
lines for each $n$ corresponds to the parameter space where the
deviation of power spectrum is smaller than 10\%. } \label{lam_m}
\end{figure}

In the left panel of Fig.~\ref{lam_m}, the relative deviation of
the power spectrum from that of the $\lcdm$ model is depicted as a
function of $\lambda$ for various total neutrino masses
with $n=2$. The right panel of Fig.~\ref{lam_m} shows the range of
total neutrino masses where the relative deviation of the power
spectrum at $k=0.174h~\Mpcin$ remains smaller than 10\%.

\begin{figure}[t]
\centering
\subfigure{\includegraphics[width=69.3mm]{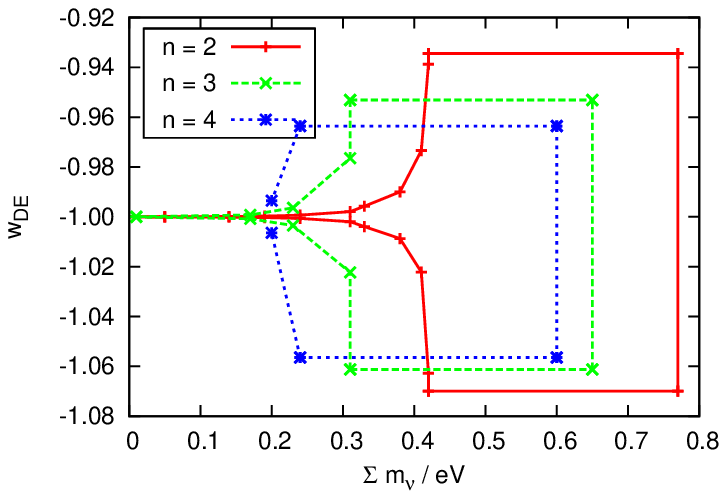}}
\subfigure{\includegraphics[width=69.3mm]{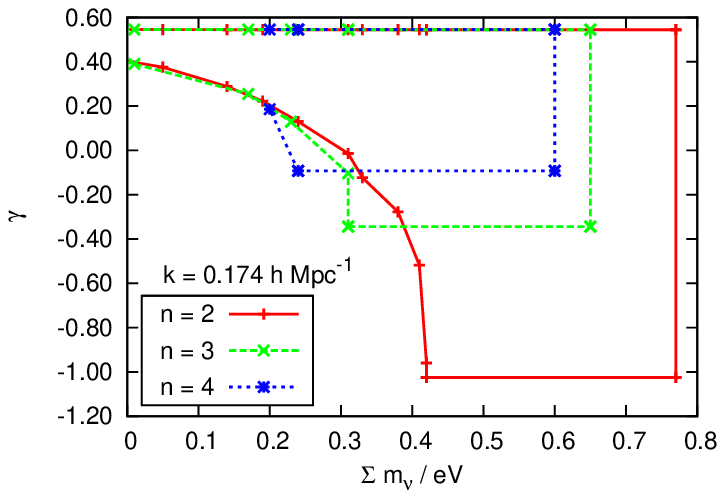}}
\caption{ Possible range of time variation of $w_\DE$ (left) and
growth index $\gamma$ (right) as a function of total neutrino
mass. $\gamma$ is measured at the comoving wave number $k=0.174
h~\Mpcin$.} \label{m_wg}
\end{figure}

As seen in the right panel, models with smaller values of
$\lambda$, which tend to exhibit a larger deviation from the $\lcdm$
model, become compatible with the observation of fluctuations. 
In other words, models with deviations of $w_\DE$ and the growth
index $\gamma(z)=\log\mk{\dot{\delta}/H\delta}/\log\Omega_m$ large
enough to be observable become viable thanks to the suppression of
small-scale fluctuations due to finite neutrino masses. 
Figure \ref{m_wg} depicts the maximum range of the time variation of
$w_\DE$ and $\gamma$ that can be realized for each value of the
total neutrino mass. That is, if we choose a minimum possible value
of $\lambda$ for each total neutrino mass, $w_\DE$
varies between the upper and lower edges of the spatula shape at
the corresponding neutrino mass. Similarly, $\gamma$ evolves
between the upper and lower edges of the ax-shape. Note that
$\gamma$ takes an almost constant value, $0.55$, in the $\lcdm$ model,
which corresponds to the upper bound in $f(R)$ gravity.

Finally, we comment on how nonlinear effects may modify our
results. So far, nonlinear effects on matter clustering in $f(R)$
gravity have been analyzed in higher order perturbation theory
\cite{Koyama:2009me} and N-body simulations \cite{Oyaizu:2008tb}.
The latter has shown that nonlinear effects lower the relative
deviations of the power spectrum, $(P_\frg - P_\lcdm)/P_\lcdm$, by
about 5\% in a specific case $f(R)=R-2\Lambda-f_{R0}R_0^2/R$ with
$|f_{R0}|=10^{-4}$, where $R_0$ is the current value of the scalar
curvature. It has also been shown that the nonlinear effects tend
to suppress the anomalous growth of small-scale fluctuations
observed in linear perturbation in $f(R)$ theory. We may therefore
conclude that nonlinear effects will not change our results
qualitatively.

In conclusion, neutrino rest-masses with
$\sum_{\nu}m_{\nu}=O(0.5~\eV)$ relax the most critical constraint
on $f(R)$ gravity, which follows from the anomalous $k$-dependent
growth of density perturbations in the cold dark matter + baryon
component at recent redshifts, making possible for $w_\DE$ and
$\gamma(k)$ deviate from those of the $\lcdm$ model noticeably.
One can distinguish the effects of $f(R)$ and massive neutrinos by 
analyzing the ratio of correlation between galaxies and curvature 
perturbation probed by weak lensing to that between galaxies and 
velocity field, namely, an estimator $E_G=\Omega_m/(F \beta)$, where 
$\beta$ is the growth rate\cite{Lombriser:2010mp}.  
According to the current observational analysis, 
both Einstein gravity and $f(R)$ gravity are consistent with 
the data.
However, in the future, one can break the degeneracy in principle.

AS acknowledges RESCEU hospitality as a visiting professor. He was
also partially supported by the grant RFBR 08-02-00923 and by the
Scientific Programme ``Astronomy'' of the Russian Academy of
Sciences. This work was supported in part by JSPS Research
Fellowships for Young Scientists (HM), JSPS Grant-in-Aid for
Scientific Research No.\ 19340054 (JY), Grant-in-Aid for
Scientific Research on Innovative Areas No. 21111006 (JY), JSPS
Core-to-Core program ``International Research Network on Dark
Energy'', and Global COE Program ``the Physical Sciences
Frontier'', MEXT, Japan.

\end{document}